\documentclass[pra,twocolumn,aps,superscriptaddress,showpacs]{revtex4-1}
\usepackage{hyperref}
\usepackage{graphicx}
\usepackage{amsmath}
\usepackage{amsfonts}
\usepackage{amssymb}
\usepackage{epsfig}
\usepackage{subfigure}
\usepackage{mathtools}
\usepackage{braket}
\usepackage[usenames,dvipsnames]{color}
\usepackage{setspace}
\usepackage[normalem]{ulem}
\usepackage{cancel}
\usepackage{times}
\usepackage{xcolor}
\hypersetup{
      colorlinks=true,
      citecolor=blue,
      linkcolor=blue,
      urlcolor=blue}

\usepackage{filemod}

\def\be{\begin{equation}}
\def\ee{\end{equation}}
\def\bg{\begin{equation}\begin{gathered}}
\def\eg{\end{gathered}\end{equation}}

\begin{document}
\title{Staggered {superfluid} phases of dipolar bosons in {two-dimensional square} lattices}
\author{Kuldeep Suthar}
\affiliation{Institute of Theoretical Physics, Jagiellonian University in Krak\'ow, \L{}ojasiewicza 11, 30-348 Krak\'ow, Poland}
\author{Rebecca Kraus}
\affiliation{Theoretical Physics, Saarland University, Campus E2.6, D--66123 Saarbr\"ucken, Germany}
\author{Hrushikesh Sable}
\affiliation{Physical Research Laboratory, Ahmedabad - 380009, Gujarat, India}
\affiliation{Indian Institute of Technology Gandhinagar, Palaj, Gandhinagar - 382355, Gujarat, India}
\author{Dilip Angom}
\affiliation{Physical Research Laboratory, Ahmedabad - 380009, Gujarat, India}
\author{Giovanna Morigi}
\affiliation{Theoretical Physics, Saarland University, Campus E2.6, D--66123 Saarbr\"ucken, Germany}
\author{Jakub Zakrzewski}
\affiliation{Institute of Theoretical Physics, Jagiellonian University in Krak\'ow, \L{}ojasiewicza 11, 30-348 Krak\'ow, Poland}
\affiliation{Mark Kac Complex Systems Research Center, Jagiellonian University in Krakow, \L{}ojasiewicza 11, 30-348 Krak\'ow, Poland}
\date{\today} 

\begin{abstract}
{We study the quantum ground state of ultracold bosons in a two-dimensional square lattice. The bosons interact via the repulsive dipolar interactions and s-wave scattering. The dynamics is described by the extended Bose-Hubbard model including correlated hopping due to the dipolar interactions, the coefficients are found from the second quantized Hamiltonian using the Wannier expansion with realistic parameters. We determine the phase diagram using the Gutzwiller ansatz  in the regime where the coefficients of the correlated hopping terms are negative and can interfere with the tunneling due to single-particle effects. We show that this interference gives rise to staggered superfluid and supersolid phases at vanishing kinetic energy, while we identify parameter regions at finite kinetic energy where the phases are incompressible. We compare the results with the phase diagram obtained with the cluster Gutzwiller approach and with the results found in one dimension using DMRG.}
\end{abstract}

\maketitle

\section{Introduction}

Ultracold atoms in optical lattices are an ideal platform to simulate complex quantum many-body systems of condensed matter
physics~\cite{lewenstein_07,bloch_08,Lewenstein12BOOK,gross_17}. One paradigmatic example is the Bose-Hubbard model~\cite{fisher_89,jaksch_98}. Here,
the possibility to tune the model's coefficients by means of external fields permits one to experimentally characterize the quantum phase transition between superfluid (SF) {to} Mott insulator (MI)~\cite{greiner_02,bakr_10}.

The recent experimental realization of ultracold dipolar gases in optical lattices \cite{depaz_13,moses_15,covey_16,baier_16,reichsollner_17,moses_17,bohn_17} opens the perspective to study the quantum properties of matter in a lattice emerging from the interplay between long-range interactions, contact interactions, and quantum fluctuations. In the theoretical model, long-range dipole-dipole interactions give rise to additional terms in the Bose-Hubbard model. These are density-density repulsive interactions, frustrating the occupation of neighbouring sites, and correlated tunneling terms, including tunneling of pairs of bosons and hopping depending on the site density \cite{sowinski_12,luhmann_12,maik_13,jurgensen_14,cartarius_17}. Similar additional interaction terms also play a key role in determining the properties of superconductors~\cite{rainer_93,hirsch_94,amadon_96}.
Most theoretical studies of the phase diagram of dipolar bosons include solely the density-density repulsive term. These works showed that the competition of this term with tunneling and contact interactions can gives rise to density modulations ~\cite{goral_02,yi_07}. The corresponding phases are denoted by charge density wave (CDW), when the phase is incompressible, and by supersolid (SS), when the phase is superfluid (for a review see \cite{Lahaye09,Baranov12}). Moreover, in one dimension the same model for unit mean density hosts a topological Haldane insulator  phase \cite{Torre06,rossini_12}.

The phases emerging from correlated hopping are relatively unexplored.  They are known in condensed matter context, where for instance 
density-dependent tunneling has being coined as bond-charge  interaction~\cite{hirsch_89}. In standard solid state materials, however, the 
role of interaction-induced tunneling is experimentally difficult to observe. In contrast, ultracold dipolar gases in optical lattices provide the unique
possibility to fully characterize the role of correlated hopping in determining the quantum phases and dynamics~\cite{jurgensen_14,baier_16}. 
Theoretical studies predict that correlated tunneling can strongly modify the phase diagram \cite{dutta_15} 
and is responsible for the appearance of
novel SF phases~\cite{sowinski_12,johnstone_19}. To mention some,  it can give rise to superfluidity with complex order parameters~\cite{jurgensen_15,luhmann_16},
or suppress the SS phase and give rise to phase-separation~\cite{maik_13}. Investigations at fractional density 3/2
predict interesting anyonic excitations \cite{Wikberg12,Duric17}. The interplay between contact and dipolar interactions were analysed in one dimension in Refs. \cite{sowinski_12,cartarius_17,Biedron18,kraus_20}
taking also care of varying the parameters according to the physical constrains of the system. The studies of Refs. \cite{Biedron18,kraus_20}, in particular, determined the ground state properties by means of density matrix renormalization group (DMRG) approach \cite{Schollwoeck11} and reported so-called staggered superfluid phases, namely, superfluid phases whose order parameter is staggered. Moreover, in Ref. \cite{kraus_20} it was predicted that correlated and single-particle hopping can interfere destructively, resulting in incompressible phases even for relatively shallow lattice, where one would otherwise expect superfluidity. 

In this work we theoretically analyse the phases of a two-dimensional lattice of ultracold dipolar bosons by means of a Bose-Hubbard model which consistently includes density-density interactions and correlated hopping. We are particularly interested in the fate of the interference between correlated and single-particle hopping in two dimensions and in a square lattice. We determine the quantum ground state using primarily the single-site standard Gutzwiller mean-field (SGMF) approach~\cite{rokshar_91,krauth_92,sheshadri_93} truncating the dipolar interactions first to nearest neighbours (NN) and then to next-nearest neighbours (NNN) terms. We then analyze the role of quantum fluctuations on the staggered phases using cluster Gutzwiller mean-field (CGMF) theory~\cite{buonsante_04,yamamoto_09,pisarski_11,mcintosh_12,yamamoto_12,luhmann_13}.

{This paper is structured as follows. In Sec.~\ref{model} we introduce the extended Bose-Hubbard model at the basis of our investigations and discuss the theoretical approaches
employed in the present work. The ground-state phase diagram at fixed average densities are presented in Sec.~\ref{pd_den}. In Sec.~\ref{pd_v}
we examine the phase diagram at fixed nearest neighbour interaction. Finally, we discuss the conclusions and outlooks in Sec.~\ref{conc}.}

\section{Model and methods}

{We consider a gas of ultracold bosons which are confined in two dimensions in a square optical lattice. The motion in the direction orthogonal to the plane is assumed to be frozen out. The bosons interact via the contact potential and the dipolar interactions. The dipoles are polarized perpendicularly with respect to the plane of the lattice, thus they interact repulsively and isotropically in the plane. In this section we introduce the extended Bose-Hubbard model at the basis of our investigation, we discuss the parameter regime on which we focus and describe the theoretical methods we employ in order to determine the quantum ground state.}

\label{model}
\subsection{Extended Bose-Hubbard model}

We assume a grand-canonical ensemble. The bosons are confined by a square lattice in the $x-y$ plane with $L=L_x\times L_y$ sites and periodic boundary conditions. We denote by $\hat b_{p,q} $ and $\hat b_{p,q} ^\dagger$ the operators annihilating and creating a particle at site $\{p,q\}$ with commutation relations $[\hat b_{p,q} ,\hat b_{p',q'}^\dagger]=\delta_{p,p'}\delta_{q,q'}$. {Here, $p$($q$) is the lattice site index along the $x$($y$) direction.} Let $\hat n_{p,q}=\hat b_{p,q} ^\dagger\hat b_{p,q}$ denote the corresponding number operator. The extended Bose-Hubbard model (eBHM) we consider is described by the Hamiltonian $ \hat{H}$ which we separate into the contribution of onsite, nearest-neighbour and next-nearest neighbour interactions:
\begin{align}
 \hat{H}=\hat{H}_{0}+\hat{H}^{(1)}+\hat{H}^{(2)}.
 \label{ham}
\end{align}
Here, $\hat{H}_0$ is the standard Bose Hubbard model:
 \begin{eqnarray}
 	\hat{H}_0 &= &-t\sum_{p,q}\left(
 	\hat{b}^{\dagger}_{p+1,q}\hat{b}_{p,q} 
 	+ \hat{b}^{\dagger}_{p,q+1}\hat{b}_{p,q} 
 	+ \rm{H.c.} \right) \nonumber\\
 	&& +\sum_{p,q} \hat{n}_{p,q} 
 	\left[ - \mu 
 	+ \frac{U}{2} (\hat{n}_{p,q} - 1) \right]\,,
 \label{BH}
 \end{eqnarray}
where $t$ is the hopping term due to single-particle effects and is isotropic, $U$ denotes the onsite interaction term, and $\mu$ is the chemical potential. 
The nearest neighbor (NN) and next-to-nearest neighbor (NNN) contributions to the Hamiltonian  contain {the terms due to both the contact and} the dipolar interactions. In detail, 
\begin{eqnarray}
  \hat{H}^{(1)} &= & V \sum_{p,q} \hat{n}_{p,q} \left(\hat{n}_{p+1,q} + \hat{n}_{p,q+1}\right) \nonumber \\
		      && - T \sum_{p,q}\bigg[\hat{b}^{\dagger}_{p+1,q} \big(\hat{n}_{p,q} 
			 + \hat{n}_{p+1,q}\big)\hat{b}_{p,q} \nonumber \\ 
		      && + \hat{b}^{\dagger}_{p,q+1} \big(\hat{n}_{p,q} 
		         + \hat{n}_{p,q+1} \big)\hat{b}_{p,q} 
			 + \rm{H.c.} \bigg] \nonumber \\
		      && + \frac{P}{2} \sum_{p,q} 
		       \left(\hat{b}^{\dagger 2}_{p+1,q} + \hat{b}^{\dagger 2}_{p,q+1} \right)\hat{b}^{2}_{p,q} + \rm{H.c.}\,, 
\label{ebhm}
\end{eqnarray}
where $V$ is the density-density interaction between neighbouring sites, $T$ is the amplitude scaling density-dependent tunneling, and $P$ is the pair tunneling coefficient. 
The Hamiltonian corresponding to the NNN coupling is specific to 2D square lattice and couples sites along the diagonals of the lattice.
Explicitly:
\begin{align}
\hat{H}^{(2)} =& V_{\rm \text{diag}}\sum_{p,q} \hat{n}_{p,q} \big(\hat{n}_{p+1,q+1} + \hat{n}_{p-1,q+1} \nonumber\\
&+ \hat{n}_{p+1,q-1} + \hat{n}_{p-1,q-1}\big) \nonumber \\
& - T_{\rm \text{diag}} \sum_{p,q}\bigg[\hat{b}^{\dagger}_{p+1,q+1} \big(\hat{n}_{p,q} 
+ \hat{n}_{p+1,q+1}\big)\hat{b}_{p,q} \nonumber \\ 
& + \hat{b}^{\dagger}_{p-1,q+1} \big(\hat{n}_{p,q} 
+ \hat{n}_{p-1,q+1} \big)\hat{b}_{p,q} + \rm{H.c.} \bigg] \,.
\label{ebhm_h2}
\end{align}
The coefficients scale the corresponding terms as in Eq. \eqref{ebhm}. The subscript ``diag" indicates that we accounted for the lattice geometry in evaluating the interactions between NNN, as we specify in the following.

\subsection{Parameters}
\label{Sec:Par}

The coefficients of the extended Bose-Hubbard model, Eq. \eqref{ham}, are here calculated by taking into account the algebraic scaling of the dipolar interactions, which is truncated at the NN or at the NNN, and the lattice geometry. Moreover, in determining the phase diagram we tune the strength of the contact interactions and of the dipolar interactions: all other coefficients are systematically scaled. The procedure extends the one implemented in Ref.~\cite{kraus_20} to a two-dimensional square lattice. We now provide some details of how we determine the coefficients, and refer the reader to Ref. \cite{kraus_20} for details. 

The potential confining the bosons in the $x-y$ plane has the form $V_{\rm latt}(\textbf r) = V_0 [\sin^{2}(kx) + \sin^{2}(ky)]$ , with $k$ the wave number and $\textbf r=(x,y)$, while the motion in the $z$ direction is frozen out.  Moreover, the bosons interact via the contact interaction, $U_{g}(\textbf r)$, and the repulsive dipolar potential $U_{\rm dip}(\textbf r)$, which decays with the distance $r$ as $r^{-3}$. The bosons are at zero temperature and the lattice depth $V_0$ is  sufficiently deep so we may limit the analysis to the lowest band. We fix $V_0=8E_R$ with the recoil energy $E_R=\hbar^2k^2/2m$ and $m$ is the particle's mass. 
The single-particle tunneling coefficient is given by the overlap integral \cite{jaksch_98}
\begin{equation}
	t_{ij} = - \int^{}_{} d\textbf{r}~w_{i}(\textbf r) 
	         \left[ -\frac{\hbar^2}{2m} \nabla^{2} + V_{\rm latt}
		 (\textbf r) \right] w_{j}(\textbf r),
\end{equation}
where $w_i({\mathbf r})$ is a product of standard (real-valued) Wannier functions at  $i\equiv \{p,q\}$. Note that due to separability of $V_{\rm latt}$ the coefficients $t_{ij}$ are non-zero only along $x$ or along the $y$ directions. Therefore, ``diagonal'' tunnelings such as e.g. $i\equiv \{p,q\}$ and $j\equiv \{p+1,q\pm1\}$ are strictly vanishing. The situation is different for the interaction terms. 

We denote the interactions by the potential $U_{\rm int}(\textbf r) = U_{g}(\textbf r) + U_{\rm dip}(\textbf r)$. The corresponding
interaction coefficients in the site basis read
\begin{equation}
	V_{ijst} = \frac{1}{2} \int d{\textbf{r}}~d{\textbf{r}'} 
			  w_i(\textbf{r})w_j(\textbf{r}')U_{\rm int}(\textbf{r} - \textbf{r}') w_s(\textbf{r}')w_t(\textbf{r}),
\end{equation}
and contribute to the interaction Hamiltonian term
\be H_{\rm int}=\sum_{ijst} V_{ijst}\hat b^\dagger_i \hat b^\dagger_j \hat b_s \hat b_t \,.
\ee 
By truncating up to the NNN terms we then obtain terms such as pair tunneling and density-dependent tunneling terms. 

In our calculations we keep the lattice depth constant. Due to our choice of $V_0$, standard approximations are justified, in particular a direct calculation shows that tunneling coefficients $t_{ij}$ between sites 
separated by more than one lattice spacing are negligible (this is not so in more shallow lattices, see e.g. \cite{Trotzky12}). We note that  the strength of the coefficients of Hamiltonian~(\ref{ebhm}) and (\ref{ebhm_h2}) follow $|V| > |T| \gg |P|$ and that we can neglect terms where $V_{ijst}$ couple four different lattice sites. Moreover,
$|V_{\rm \text{diag}} |> |T_{\rm \text{diag}}|$. In our isotropic lattice geometry the overlap integral giving $P$ is orders of magnitude smaller than $V$ or $T$, moreover, it is much smaller than the coefficients appearing in $ \hat{H}^{(2)}$. We have checked explicitly that taking it into account  does not modify our results.
For that reason, from now on we omit the term proportional to $P$ in \eqref{ebhm} - this justifies also that  the corresponding term is also missing in \eqref{ebhm_h2} from the very beginning. For completeness, we remark that the pair tunneling coefficient can be enhanced by modifying the confinement in the $z$ direction \cite{sowinski_12}.

Let us repeat that
{in our calculations we vary the parameters by keeping the lattice depth constant and tuning both the contact and the dipolar interactions}. While the {contact interactions are tuned by means of} Feshbach resonances \cite{Chin10}, {dipolar forces can be varied by rotating the 
orientation of dipoles as proposed by \cite{Giovanazzi02} and demonstrated in \cite{baier_16,Tang18}. A similar effect may be also} obtained - within the limitations mentioned above -  by changing the depth of the optical lattice and the contact potential keeping the strength of the dipolar interaction fixed.  

In Ref.~\cite{kraus_20} we analysed the parameter regime where $T<0$. In particular, mean-field considerations predict that the density-dependent tunneling term scaling with $T$ can interfere destructively with the single-particle hopping scaling with $t$. Destructive interference is found when the condition 
\begin{equation}
\label{T:t}t=|T|(2\rho-1)
\end{equation} is fulfilled at average densities $\rho>1/2$ \cite{kraus_20}. In Ref. \cite{kraus_20} we verified that this interference cuts the phase diagram into two topologically different superfluid phases. In particular, there where correlated hopping dominates the SF order parameter is staggered. This is the regime we analyse in this work.


\subsection{Gutzwiller mean-field theory}
\label{gucio}

We study the ground state properties of the model by means of the so called site-decoupled Gutzwiller mean-field or SGMF 
approach~\cite{rokshar_91,krauth_92,sheshadri_93}. In its standard version the bosonic annihilation (creation) 
operator is decomposed as $$\hat{b}_{p,q} = \langle \hat{b}_{p,q} \rangle + \delta \hat{b}_{p,q}\,,$$ where 
$$\langle \hat{b}_{p,q} \rangle \equiv \phi_{p,q}$$ is the SF order parameter, which signals long-range phase coherence, and 
$\delta \hat{b}_{p,q}$ is the fluctuation operator. In the presence of density-dependent tunneling and pair tunneling this approach has to be generalized
including additional order parameters connected with density{-dependent}  and pair hopping \cite{johnstone_19}, which we define below.

The many-body wave function is given by the Gutzwiller ansatz
\begin{equation}
 |\Psi_{\rm GW}\rangle = \prod_{p,q}|\psi_{p,q}\rangle =\prod_{p,q}
		       \sum_{n=0}^{n_{\text{max}}} c^{(p,q)}_n|n\rangle_{p,q},
 \label{gw}
\end{equation}	
where ${|n\rangle_{p,q}}$ are the occupation number basis states, 
$n_{\text{max}}$ is the maximum 
number of bosons at each site, and $c^{(p,q)}_n$ are basis expansion coefficients 
of  $|\psi_{p,q}\rangle$.  The state $|\Psi_{\rm GW}\rangle$ is normalized to unity by imposing $\sum |c^{(p,q)}_n|^2=1$.
Using Eq. \eqref{gw} the SF order parameter reads
\begin{equation}
  \phi_{p,q} = \langle\Psi_{\rm GW}|\hat{b}_{p,q}|\Psi_{\rm GW}\rangle=\sum_{n} \sqrt{n}~{c^{(p,q)}_{n-1}}^{*}c_n^{(p,q)}.
\end{equation}
It is finite 
for the SF, SS and their respective staggered phases whereas it is zero for incompressible 
phases. 

The average density $\rho$ is given by $$\rho=\sum_{p,q} n_{p,q}/L$$ where $L=L_x\times L_y$ is the system size and 
$n_{p,q}$ is the density at $(p,q)$th site: 
\begin{equation}
  n_{p,q} = \langle\Psi_{\rm GW}| \hat{n}_{p,q}
            |\Psi_{\rm GW}\rangle 
          = \sum_{n} n |c_n^{(p,q)}|^2.
\end{equation}
Finally, we introduce the density-assisted correlation order parameter, which reads 
\be
  \eta_{p,q} = \langle\Psi_{\rm GW}|\hat{n}_{p,q}\hat{b}_{p,q} |\Psi_{\rm GW}\rangle
	      = \sum_{n} \sqrt{n} (n-1)~{c^{(p,q)}_{n-1}}^{*}c_n^{(p,q)}\,.
 \label{eta_chi}
\ee
Its behaviour allows us to identify staggered phases \cite{johnstone_19}. 

Using these definitions the mean-field Hamiltonian of the system may be written as a sum  
 of  single-site Hamiltonians $\hat{H}_{\rm MF} =\sum_{p,q} \hat{h}_{p,q}$. For NN interactions the individual summands $\hat{h}_{p,q}$  read
\begin{widetext}
  \begin{eqnarray}
    \hat{h}_{p,q}
            &=& -t\Big[\left(\phi^{*}_{p+1,q}\hat{b}_{p,q} 
	      + \hat{b}^{\dagger}_{p,q}\phi_{p-1,q} - \phi^{*}_{p+1,q}\phi_{p,q}\right)
	      + \left(\phi^{*}_{p,q+1}\hat{b}_{p,q} + \hat{b}^{\dagger}_{p,q}\phi_{p,q-1}
	      - \phi^{*}_{p,q+1}\phi_{p,q}\right) + {\rm H.c.}\Big]
              \nonumber\\
	    &&+ \frac{U}{2}\hat{n}_{p,q}(\hat{n}_{p,q}-1)
	      - \mu\hat{n}_{p,q}
	      + V \Big[\hat{n}_{p,q}\Big(\langle\hat{n}_{p+1,q}\rangle
              + \langle\hat{n}_{p,q+1}\rangle + \langle\hat{n}_{p-1,q}\rangle + \langle\hat{n}_{p,q-1}\rangle \Big) 
	      - \langle \hat{n}_{p,q} \rangle
             \Big(\langle\hat{n}_{p+1,q}\rangle 
             + \langle\hat{n}_{p,q+1}\rangle \Big)\Big]
              \nonumber\\
	   &&- T \Big[\Big(\hat{b}^{\dagger}_{p,q} \eta_{p-1,q} + \phi^{*}_{p+1,q} \hat{n}_{p,q}\hat{b}_{p,q}
	     + \eta^{*}_{p+1,q} \hat{b}_{p,q} + \hat{b}^{\dagger}_{p,q} \hat{n}_{p,q} \phi_{p-1,q} 
	     - \phi^{*}_{p+1,q} \eta_{p,q} - \eta^{*}_{p+1,q} \phi_{p,q} \Big)  
              \nonumber\\
	   &&+ \Big(\hat{b}^{\dagger}_{p,q} \eta_{p,q-1} + \phi^{*}_{p,q+1} \hat{n}_{p,q}\hat{b}_{p,q} 
	     + \eta^{*}_{p,q+1} \hat{b}_{p,q} + \hat{b}^{\dagger}_{p,q} \hat{n}_{p,q} \phi_{p,q-1}
	     - \phi^{*}_{p,q+1} \eta_{p,q} - \eta^{*}_{p,q+1}\phi_{p,q} \Big) 
             + {\rm H.c.}           
             \Big], 
\label{ebhm_mf}
\end{eqnarray}
\end{widetext} 
The mean field Hamiltonian corresponding to higher order (NNN) terms is found by applying the same procedure and has a similar form. 
This form makes evident that density-assisted tunneling and single-particle hopping can interfere when both the SF order parameter and $ \eta_{p,q}$ are different from zero. 

We solve the model by diagonalizing the single-site Hamiltonians coupled through the mean-field 
self-consistently~\cite{jaksch_98,goral_02,zakrzewski_05,scarola_06,menotti_07,iskin_11,kuno_17,danshita_17,
bai_18,pal_19,bandyopadhyay_19,suthar_19,johnstone_19}.
In order to solve the on-site Hamiltonian~[Eq.(\ref{ebhm_mf})], we initialize the Gutzwiller 
coefficients by $1/\sqrt{n_{\rm max}}$ on each-site and then {evaluate the corresponding initial order parameters}. {We then sequentially diagonalize the local Hamiltonians, the order parameters
are redefined using the ground states found from these diagonalizations}. With each diagonalization, the order parameters $\phi$,
and $\eta$'s are updated and this procedure is repeated until the convergence 
criteria of the order parameter is satisfied. For the present work, we consider the 
convergence criterion of $10^{-12}$ in $\phi$ for two consecutive iterations.  
The ground-state of the 
system is obtained by checking the initial order parameters with uniform {density} and {with} density wave orders, 
and by then analyzing the energy of the system.

\subsection{Cluster Gutzwiller mean-field theory}
In the SGMF theory, the inter-site coupling is
incorporated through the mean fields. For instance, the nearest neighbour 
hopping is through the SF order parameter. This accounts for the poor 
resolution of  inter-site correlations. This shortcoming can be partially overcome 
by the application of the CGMF method. In CGMF one make partitions of the entire system, say 
$L_{x} \times L_{y}$ lattice, into $W$ clusters of dimension $M \times N$.
Here,  $W = (L_{x} \times L_{y})/(M \times N)$ is an integer, and the system 
Hamiltonian is written as a sum of the cluster Hamiltonians. A cluster
Hamiltonian has two types of terms. First, the intra-cluster terms involving 
lattice sites from within the cluster only. 
And, second, the inter-cluster terms which couple the lattice sites at the boundary 
of the cluster with the sites of the neighbouring clusters.
Only the latter are treated in the mean field level as in SGMF. The
detailed implementation of the CGMF can be found in \cite{buonsante_04,
yamamoto_09,pisarski_11,mcintosh_12,yamamoto_12,luhmann_13,deng_15,sachdeva_17,bai_18,
suthar_19}. The Hamiltonian of a single cluster is 
\allowdisplaybreaks
\begin{widetext}
\vspace{-1em}
\begin{eqnarray}
   \hat{H}_C = &-& t \sum_{p,q}^{'} \left(\hat{b}_{p+1,q}^{\dagger}\hat{b}_{p,q} + 
                   \hat{b}_{p,q+1}^{\dagger}\hat{b}_{p,q} + {\rm H.c.} \right)
                 - t \sum_{p,q \in \delta C}\big(\phi_{p+1,q}^{*}\hat{b}_{p,q} +
                   \hat{b}_{p,q}^{\dagger}\phi_{p-1,q}  
                 + \phi_{p,q+1}^{*}\hat{b}_{p,q} + 
                   \hat{b}_{p,q}^{\dagger}\phi_{p,q-1} + {\rm H.c.}\big)\nonumber \\
                &+&\frac{U}{2}\sum_{p,q} \hat{n}_{p,q}(\hat{n}_{p,q}-1) -
                   \mu\sum_{p,q}\hat{n}_{p,q} 
		 + V \sum_{p,q}^{'} \hat{n}_{p,q}\Big(\hat{n}_{p+1,q} + 
                   \hat{n}_{p,q+1}\Big) \nonumber \\ 
	        &+&V \sum_{p,q \in \delta C}\hat{n}_{p,q}\Big(
                   \langle \hat{n}_{p+1,q} \rangle + \langle \hat{n}_{p,q+1} 
		   \rangle + \langle \hat{n}_{p-1,q} \rangle + \langle \hat{n}_{p,q-1} \rangle \Big) \nonumber \\
	        &-&T \sum_{p,q}^{'} \Big(\hat{b}_{p+1,q}^{\dagger}
                   (\hat{n}_{p,q} + \hat{n}_{p+1,q}) \hat{b}_{p,q} 
                 + \hat{b}_{p,q+1}^{\dagger}(\hat{n}_{p,q} + \hat{n}_{p,q+1})
                   \hat{b}_{p,q} + {\rm H.c.} \Big) \nonumber \\ 
	        &-&T \sum_{p,q \in \delta C}\Big[\Big(\hat{b}^{\dagger}_{p,q}\eta_{p-1,q} + 
                   \phi^{*}_{p+1,q}\hat{n}_{p,q}\hat{b}_{p,q} + \eta^{*}_{p+1,q}\hat{b}_{p,q} 
		 + \hat{b}_{p,q}^{\dagger}\hat{n}_{p,q}\phi_{p-1,q} \Big)\nonumber \\
	        &+&\Big(\hat{b}^{\dagger}_{p,q}\eta_{p,q-1} + \phi^{*}_{p,q+1} \hat{n}_{p,q}\hat{b}_{p,q} 
	         + \eta^{*}_{p,q+1}\hat{b}_{p,q} + \hat{b}_{p,q}^{\dagger}\hat{n}_{p,q}\phi_{p,q-1} \Big) 
		 + {\rm H.c.}\Big].
  \label{clus_hamil}
\end{eqnarray}
\end{widetext}
The prime in the summation of the intra-cluster terms is to 
indicate that the lattice sites $(p+1,q)$ and $(p,q+1)$ are also within the 
cluster. Here, $\delta C$ is the set of lattice sites at the boundary of the 
cluster.

The ground state of the cluster Hamiltonian is 
\begin{equation}
  \Ket{\Psi_c} = \sum_{l}C_l \Ket{\Phi_c}_\ell\,,
  \label{cluster_gs}
\end{equation}
where
\begin{equation}
   \ket{\Phi_c}_\ell = \prod_{q=0}^{N-1}\prod_{p=0}^{M-1} \ket{n}_{p,q}\,.
\end{equation}
where $\ket{n}_{p,q}$ is the occupation number basis at the $(p,q)$ lattice
site of the cluster, and $\ell \equiv \{n_{00}, n_{10}, \ldots, n_{M-1,0}, n_{01}, n_{11},\ldots
n_{M-1,1}, \ldots, n_{M-1,N-1}\}$ is the index quantum number to identify
the cluster state. The ground state of the entire system, as in the SGMF, is 
the direct product of the cluster ground states
\begin{equation}
  \ket{\Psi^c_{\rm GW}} = \prod_k\ket{\Psi_c}_k,
\label{cgw_state}
\end{equation}
where $k$ is the cluster index and varies from 1 to $W$.
To solve the cluster Hamiltonian Eq.(\ref{clus_hamil}), we 
initialize the order parameters in our model. We then diagonalize the cluster
Hamiltonians, and then update the values of the order parameters based on 
the ground states obtained during the diagonalization. As in SGMF, this 
procedure is repeated till {it converges}.

\subsection{Observables}
\label{Sec:Obs}

To examine the ground-state properties of the system, we compute the 
single-particle correlation 
\begin{eqnarray} 
  M_{b} (\textbf{k}) &=& \frac{1}{L^2} \sum_{j,j'} 
			  e^{i \textbf{k}\cdot(\textbf{r}_ j - \textbf{r}_{j'})}
		          \langle \hat{b}^{\dagger}_{j} \hat{b}_{j'} \rangle
\end{eqnarray} 
as well as 
the density-density correlation or the structure factor  defined as   
\begin{equation} 
  S (\textbf{k}) = \frac{1}{L^2} \sum_{j,j'} 
	            e^{i \textbf{k}\cdot(\textbf{r}_j - \textbf{r}_{j'})}
		    \langle \hat{n}_{j} \hat{n}_{j'} \rangle.
\end{equation}
In the above definitions, $\textbf{k} = (k_x, k_y)$ is the dimensionless wave-vector and $\textbf{r}_ j\equiv (p,q)$, $\textbf{r}_{j'} \equiv (p', q')$.  Finite values of $M_b{({\bf k})}$ and $S(\pi,\pi)$ correspond to off-diagonal long range order and diagonal long range 
order, respectively~\cite{otterlo_94,batrouni_00,hebert_01}. 

Using the Gutzwiller wave function given by Eq.(\ref{gw}), $M_b(\textbf{k})$ 
 and $S(\textbf{k})$ can be rewritten as 
\begin{eqnarray} 
  M_{b} (\textbf{k}) &=& \frac{1}{L^2} \sum_{j,j'} 
                         e^{i \textbf{k}\cdot(\textbf{r}_ j - \textbf{r}_{j'})}
                         \phi^{*}_{j} \phi_{j'} \nonumber \\
  S (\textbf{k})     &=& \frac{1}{L^2} \sum_{j,j'} 
	                  e^{i \textbf{k}\cdot(\textbf{r}_j - \textbf{r}_{j'})}
		          n_{j} n_{j'}. \nonumber
\end{eqnarray}
 In the CGMF theory, the correlated wavefunction in 
Eq.(\ref{cluster_gs}) and Eq.(\ref{cgw_state}) prevents the replacement of 
the operators by their expectation values. Instead, the observables with 
operators from within a single cluster are calculated using the
cluster wavefunction defined in Eq.(\ref{cluster_gs}) and the total 
wave function (\ref{cgw_state}) is used to calculate the observables with
operators from different clusters.

The classification of various quantum phases can be obtained using the behavior of 
correlation and structure factor at zero and finite momentum $\textbf{k}$. When truncating the Hamiltonian to NN, the effect of the interactions is to give rise to spatially periodic structures with $2\times 2$ cells. Periodic density modulations, in particular, can be characterized by their sublattice distributions $(n_a,n_b)$ with $a\equiv (p,q)$ and $b\equiv (p',q')$, where $(p',q')$ is the nearest neighbor lattice site of $(p,q)$. Similarly, correlated tunneling gives rise to periodic modulation of the SF order parameter of the phase~\cite{johnstone_19}. The onset of these structures is signalled by the finite values of $S$ and $M_b$, respectively, at the wave vector $\textbf{k}=(\pi,\pi)$. 

The inclusion of NNN terms break this symmetry. In general, it is known that the NNN repulsion tends to stabilize striped order of supersolid phase. The coexistence of NN and NNN stabilizes various solid orders, for $V \geqslant 2V_{\rm diag} (V < 2V_{\rm diag})$, checkerboard (stripe) solid ordered state is formed~\cite{batrouni_95,scalettar_95,yamamoto_12}. Here, in order to distinguish between various orders of solid, supersolid and staggered phases, 
we determine the Fourier transform of the single-particle correlations $M_b(\mathbf{k})$ and {the} structure factor $S(\mathbf{k})$ at $\textbf{k} = (\pi,\pi), (0,\pi)$ and $(\pi,0)$. A finite value of those observables at $(\pi,\pi)$ reflects checkerboard order whereas a finite value at $(0,\pi)$ or $(\pi,0)$ shows 
striped order. {Correspondingly,} a finite value of $M_b(\mathbf{k})$ and $S(\mathbf{k})$ at $\textbf{k} = (\pi,\pi), (0,\pi)$ and $(\pi,0)$ is the characteristic property of a quarter-filled ordered superfluid phase~\cite{chen_08,ng_10}. Depending on the ratio between $M_b(0,0)$ and $M_b(\pi,\pi)$ the quarter-filled order superfluid phase is either a staggered or a normal superfluid. In this work we will find quarter-filled ordered staggered SS phases, which we will label by the acronym QF-SSS. Table~\ref{tab_op} summarizes the properties of different quantum phases in terms of the correlation and  order parameters.

\begin{table*}
  \caption{Classification of phases: Sublattice values of the occupation $\boldsymbol{n}=(n_a,n_b)$, single-particle correlations, and structure form factor. When single-particle correlations and structure form factor do not vanish, we report the value at which $\textbf{k}= \textbf{k}_{\rm max}$ at which their absolute value is maximum and which identify the phase. The details of the QF-SSS phase are discussed in Sec. \ref{Sec:Obs}. See also Ref. \cite{johnstone_19}.}
  \centering
  \begin{tabular}{l c c c c c c c c } \\
    \hline\hline \\
      Phase &  Acronym & $ {\boldsymbol{n}} $  &  $ M_b(\textbf{k})$ &  $ S(\pi,\pi) $\\ \\ [0.2ex]
	    \hline\hline \\		
      Mott Insulator             & MI           & $ (n,n) $           & $ 0 $                & $ 0 $ \\
      Charge Density Wave        & CDW          & $ (n_a,n_b) $ &  $ 0 $        & $\neq 0$ \\ 
      Superfluid                 & SF      & $ (n,n) $     &  $ \textbf{k}_{\rm max}=(0,0)$  & $ 0 $\\
      Supersolid                 & SS  & $ (n_a,n_b) $  &  $ \textbf{k}_{\rm max}=(0,0)$ & $ \neq 0$ \\
      Staggered Superfluid       & SSF     & $ (n,n) $    & $ \textbf{k}_{\rm max}=(\pi,\pi)$        & $0$\\
      Staggered Supersolid       & SSS & $ (n_a,n_b) $ & $ \textbf{k}_{\rm max}=(\pi,\pi)$          & $\neq 0$\\
      Quarter-Filled Staggered Supersolid & QF-SSS & other  & other  & other \\[1.5ex]
		
      \hline
  \end{tabular}
\label{tab_op}
\end{table*}

Finally, for convenience in the following we also use the average SF order parameter 
\begin{equation}
\label{phi:av}
\phi_{\rm avg} = \sum_{p,q} |\phi_{p,q}|/L\,,
\end{equation}
as well as the average  density-correlated order parameter 
\begin{align}
		\eta_{\rm avg}=\sum_{p,q}|\eta_{p,q}|/L \, . \label{average_eta}
\end{align}
{ These two order parameters serve to define quantum phases of eBHM. Here, we consider the absolute values of 
$\phi$'s and $\eta$'s as for staggered phases their distributions over the lattice sites alternate in sign and in particular 
for staggered superfluidity these observables become zero otherwise.
}

\section{Phase transitions at fixed density}
\label{pd_den}

We numerically determine the ground states of the system in the grand-canonical ensemble and at fixed density. 
For this purpose we first obtain the {values} of the chemical potential $\mu$ corresponding to a constant average density. The model parameters are 
obtained using Wannier function formalism as described in Sec. \ref{Sec:Par}, where {we change $T$ and $V$ by varying} the 
interaction strength, see Ref. \cite{kraus_20} for details. Below we choose to plot $V$ instead of the interaction strength. We remind the reader that finite values of $V$ {also imply} finite values of $T$.

The phases are identified by analyzing the behaviour of correlations and order parameters, see Sec. \ref{Sec:Obs}. We first discuss the role 
of NN interaction and density-dependent tunneling. We then investigate the role of NNN terms in determining the ground state properties.  
Note that below we label the coordination number by the parameter $z$. In our lattice geometry $z=4$.


\subsection{Phase diagram for nearest-neighbor interactions}
\begin{figure}
  \includegraphics[width=\linewidth]{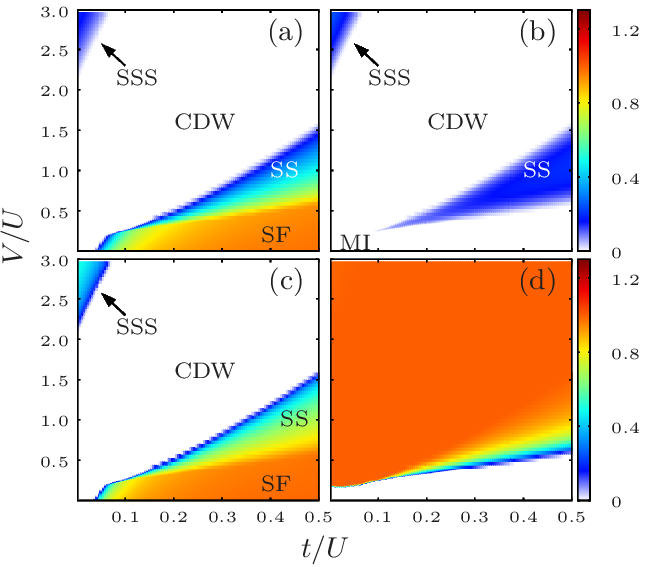} 
  \caption{Observables for the ground-state of the 2D eBHM at $\rho = 1$ for the mean-field Hamiltonian, Eq.\,\eqref{ebhm_mf}, truncated at NN. Here the ground state was calculated by means of the SGMF approach. 
	   	Subplots (a) and (b) show the Fourier transform of the off-diagonal single-particle correlation $M_b$ at $(k_x,k_y) = (0,0)$ 
		and 
	       $(k_x,k_y) = (\pi,\pi)$, respectively. The transition from MI-SF is identified by a finite value of $M_b(0,0)\geqslant M_b(\pi,\pi)$. In contrast the staggered superfluid phases correspond to the regions where $M_b(0,0)< M_b(\pi,\pi)$. Subplot (c) shows the average SF order parameter, Eq. \eqref{phi:av}. The structure factor $S(\pi,\pi)$ is shown in (d). Its finite value signals density modulated phases.
	  The system size is $L=12\times 12$, periodic boundary conditions are assumed and the maximum occupancy
	  per lattice site $n_{\rm max}$ is taken to be $8$.}
  \label{corr1}
\end{figure}

\begin{figure}
  \includegraphics[width=\linewidth]{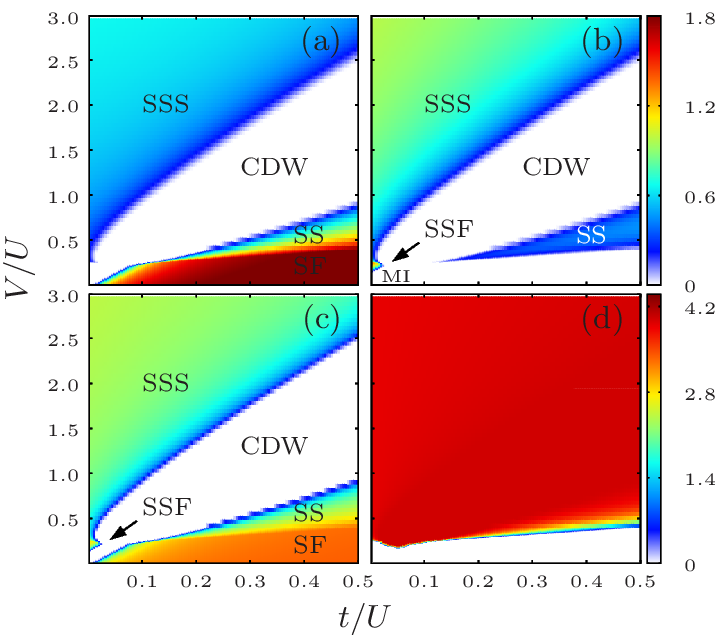} 
  \caption{Observables for the ground-state of the 2D eBHM at $\rho = 2$ for the mean-field Hamiltonian, Eq.\,\eqref{ebhm_mf}, truncated at NN. Here the ground state was calculated by means of the SGMF approach. 
		Subplots (a) and (b) show the Fourier transform of the off-diagonal single-particle correlation $M_b$ at $(k_x,k_y) = (0,0)$ and 
		$(k_x,k_y) = (\pi,\pi)$, respectively. Subplot (c) displays the average SF order parameter, Eq. \eqref{phi:av}. The structure factor $S$ at 
		$(k_x,k_y) = (\pi,\pi)$ is shown in (d). As for $\rho=1$, we consider $12\times12$ system with periodic boundary conditions
		 and $n_{\rm max}=8$.}

\label{corr2}
\end{figure}


\subsubsection{Average density $\rho = 1$}

We now discuss the phase diagram at density $\rho=1$. {We first recall that  the properties of the ground state are very 
well studied in the absence of density-dependent tunneling (for $T=0$). In this limit the inclusion of NN density-density interaction to BHM leads to additional quantum phases which we shortly review {in the following}~\cite{kuhner_00,rossini_12,Lahaye09,Baranov12}. For large $V$ bosons occupy every alternate lattice site forming CDW phase, {an incompressible phase} which spontaneously breaks sublattice symmetry. 
{When the NN interaction $V$ is} comparable to the on-site repulsion, the system breaks both $U(1)$ gauge symmetry and translational invariance to form a SS 
phase. The SS phase of eBHM is found to be stable in wide range of interaction strengths~\cite{goral_02,kovrizhin_05,menotti_07,danshita_09}.} 
{For unit filling or density the stability of SS of soft-core bosons has been demonstrated in Refs.~\cite{sengupta_05,ohgoe_12}. 
When {$4V \ll U$}, the system exhibits MI-SF phase transition as $t$ increases.} At stronger NN interactions 
{$4V\approx U$}, the MI phase becomes unstable and CDW phase replaces the MI 
phase~\cite{iskin_11,kovrizhin_05,yi_07,ohgoe_12}. Numerical studies of one dimensional lattices reveal the existence of topological Haldane phase separating 
MI and CDW phases~\cite{rossini_12,gremaud_17,sugimoto_19}. {The dependence of the hopping parameter $t$ on various phase transitions 
for fixed densities has been discussed~\cite{kimura_11}. We report the corresponding phase diagram with $T=0$ in the Appendix.}

The effect of correlated hopping in determining the phases is now visible in Fig.~\ref{corr1}, which displays the correlation $M_b$ at $(0,0)$ and $(\pi,\pi)$, the average SF order parameter, 
and the structure form factor $S(\pi,\pi)$. These quantities are reported in the  $t-V$ plane and have been obtained using the SGMF approach. 
The properties of $M_b$ and $S(\pi,\pi)$ allow us to identify the phases. At {$V/U\leqslant 0.25$}, as $t$ increases, $M_b(0,0)$ remains nonzero 
whereas it vanishes at $(\pi,\pi)$. In addition, $S(\pi,\pi)$ is zero in this region. The region where both $M_b(0,0)$ and $S(\pi,\pi)$ vanishes is MI, whereas the region with finite $M_b(0,0)$ is SF. At lower $V$ we recover the MI-SF transition of BHM~\cite{fisher_89,jaksch_98}. 
 {The critical hopping strength of MI-SF transition $t_c$ increases for {$0\leqslant 4V \leqslant U$} with finite $T$.} This shift is due to the finite value of correlated hopping, which interferes destructively with single-particle tunneling.  

At {$V/U\simeq 0.25$} we observe the onset of periodic density modulations, which are signalled by the finite value of  $S(\pi,\pi)$. The CDW phase corresponds to the large region with vanishing off-diagonal order, $M_b({\bf k})=0$. This region is centered about the line of perfect interference $|T|=t$, Eq. \eqref{T:t}. There is a direct MI-CDW transition for a finite range of values $t/U$ at {$V/U\simeq 0.25$}, which is characterized by a fast increase of $S(\pi,\pi)$ from zero to the maximum value. In the presence of SF this increase is gradual and characterised by the appearance of a finite value of $M_b(\pi,\pi)$. We identify this region with the SS phase since $|M_b(\pi,\pi)| < |M_b(0,0)|$, see Table~\ref{tab_op}. The boundary $t_c/U$ of the transition CDW-SS increases with $V$. We have checked that $t_c/U$ is shifted to larger values with respect to the phase boundary one obtains by setting $T=0$, see Appendix. The resulting domain of the CDW phase is larger. We attribute this effect to the destructive interference between single-particle and correlated hopping. 

At larger values {$V \approx 2U$} and for vanishing $t/U$ we observe the appearance of finite off-diagonal long-range order with $|M_b(\pi,\pi)| > |M_b(0,0)|$, which we identify with a staggered supersolid (SSS). This phase is due to correlated hopping, which becomes dominant at large ratios $V/U$. 

These behaviours can be understood on the basis of mean field considerations: correlated hopping favours staggered SF for $t/U\to 0$. The size of this region increases as $V$ (and thus the strength of the dipolar interactions) increases. At large ratios $t/U$, instead, the kinetic energy dominates and the phase is SF. The two contributions interfere destructively in an intermediate region, where the phase is incompressible. Diagonal long-range order is found for {$V/U\geqslant 0.25$}. We observe that the size of the SSS region at {$V/U\geqslant 0.25$} is now significantly smaller.


\subsubsection{Average density $\rho = 2$}

Figure~\ref{corr2} displays the phase diagrams of the relevant quantities in $t-V$ plane for the average density $\rho=2$. 
Similarly to the unit filling case, for {$V/U<0.25$} the phase is either MI or SF. The phase boundary $t_c$ separating the two phases depends on the strength of the dipolar interactions due to the interference between correlated hopping and single-particle hopping. For {$V/U\gtrsim 0.25$} the structure form factor $S(\pi,\pi$ is different from zero and signals the onset of density modulations. As compared to quantum phases at unit filling [Fig.~\ref{corr1}], the size of the region of insulating CDW phase is reduced whereas the size of the SSS parameter region is significantly larger. A striking difference with respect to the phase diagram at unit filling is the appearance of SSF at {$V/U\sim 0.25$} and $t/U\to 0$. 

\begin{figure}
 \includegraphics[width=\linewidth]{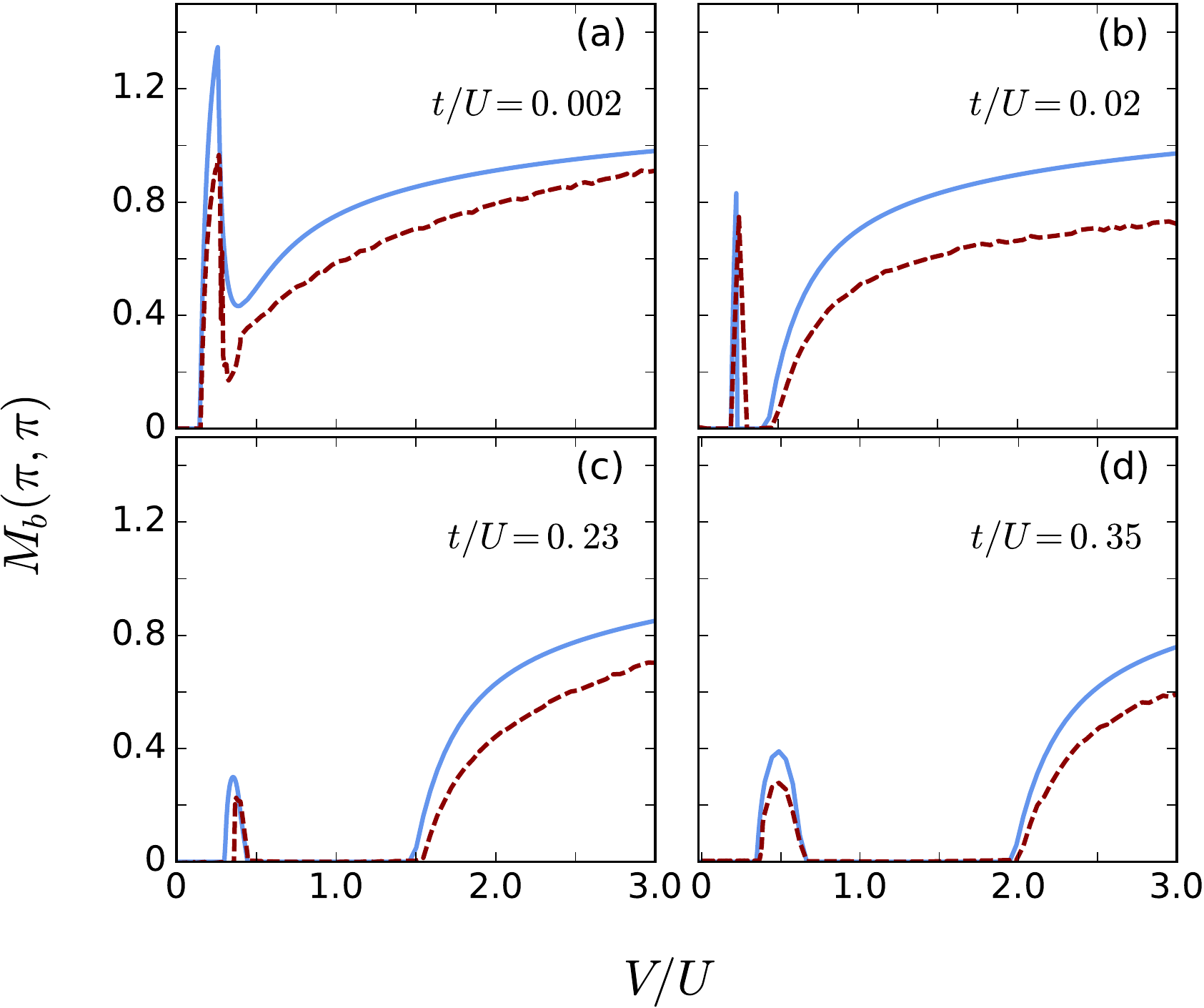}
 \caption{Off-diagonal single-particle correlation $M_{b}(\pi,\pi)$ as a 
	  function of $V/U$. The blue lines report the behaviour corresponding to Fig. \ref{corr2}(b) for hopping strengths $t/U=0.002,0.02,0.23,0.35$. The values of $t/U$ is reported in the subplots. The dashed red line is obtained using $2 \times 2$ cluster in CGMF theory. Here, we consider the maximum 
occupancy per lattice site $n_{\text{max}} = 8$: this choice of 
$n_{\text{max}}$ is sufficient to obtain converging results for densities considered.
}
 \label{corr_vs_t}
\end{figure}

The behaviour reported in Fig. \ref{corr2} qualitatively agrees with the phase diagram calculated for the same model but in one dimension and using DMRG~\cite{kraus_20}. 
{ However, it is worth noting that the parameter regime of SSF is shifted to lower $V$ values as compared to 1D.
This is due to the larger coordination number of square lattices and for $d$-dimensional lattices the staggered superfluidity 
exist at and around $2dV\approx U$, where $2d=z$ is the lattice coordination number.}
As in \cite{kraus_20}, the finite value of $T$ is responsible for the appearance of SSF at {$V/U\sim 0.25$} and $t/U\to 0$. The size of the SSF phase is now significantly smaller. In order to perform a systematic comparison with the one dimensional case, we now consider a small, fixed value $t/U$ and analyse the phases as a function of $V/U$. Figure \ref{corr_vs_t} displays the behaviour of $M_b(\pi,\pi)$ as a function of $V/U$ for (a) $t/U=0.002$, (b) $t/U=0.02$, (c) $t/U=0.23$, and (d) $t/U=0.35$. For  $t/U=0.002$ the phase is first MI. In the atomic limit, at {$V/U=0.25$} MI and several CDW phases are degenerate. Here, the phase becomes SSF due to the prevailing role of correlated hopping and is signalled by the peak of $M_b(\pi,\pi)$ at {$V/U\sim 0.25$}. This peak was also observed in one dimension and exhibits features of a continuous phase transition \cite{kraus_20}. At higher $V$ the value of $M_b$ increases again with $V$, together with the finite value of the structure form factor we identify this phase as SSS. At $t/U=0.02$, the phases SSF and SSS are separated by an incompressible CDW phase.
At higher $t$ in (c), single-particle hopping dominates over correlated tunneling. Here, the system is driven first from  SF to SS and then  to SSS with an intermediate insulating CDW phase. The width of CDW phase in between the SS and SSS phases increases and the region of SSS phase reduces with $t$. This is evident from the behaviour of $M_b$ at $t/U=0.35$ shown in Fig.~\ref{corr_vs_t}(d). 

In order to examine the effects of quantum fluctuations within our mean-field ansatz, we use the CGMF method and compute the $M_b(\pi,\pi)$ with a $2\times 2$ 
cluster. The resulting behaviour is reported by the dashed red line in Fig.\ref{corr_vs_t}. At $t/U = 0.002$ the domain of the SSF phase, signalled by 
the peak in the $M_{b}(\pi,\pi)$, is the same as in the SGMF method, while the SSF-SSS transition becomes sharper. For $t/U = 0.02$, Fig.\ref{corr_vs_t} (b), the critical value of $V/U$ 
for the CDW-SSS transition shows a small increase as compared to SGMF transition. { When $t/U$ is increased to
$t/U = 0.23$, Fig.\ref{corr_vs_t}(c), we observe a shift in the peak of $M_b(\pi,\pi)$, signalling the SS phase, to higher values of $V/U$. 
In detail, the peak occurs at $V/U = 0.3$ with SGMF, this shifts to $V/U = 0.36$ with the $2\times 2$ cluster. We note that the shift between 
the boundary SF-SS predicted by CGMF and the one predicted by SGMF is consistent with the findings reported in \cite{ohgoe_12,suthar_19} for 
the SF-SS transition. }


\subsection{Phase diagram for next nearest-neighbor interactions}
\begin{figure}
  \includegraphics[width=\linewidth]{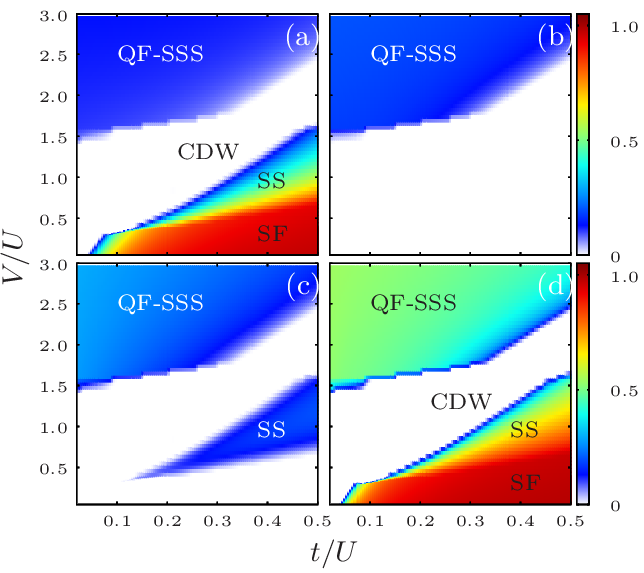}
  \caption{Observables for the ground-state of the 2D eBHM at $\rho = 1$ for the mean-field Hamiltonian, Eq.\,\eqref{ebhm_mf}, truncated at NNN, and calculated by means of the SGMF approach. 
	   	Subplots (a), (b) and (c) show the Fourier transform of the off-diagonal single-particle correlation $M_b$ at $(k_x,k_y) = (0,0)$, $(0,\pi)$, and 
	   	$(k_x,k_y) = (\pi,\pi)$, respectively. At $V/U \gtrsim 1.5$, the finite values of $M_b$ both at $(0,\pi)$ and $(\pi,\pi)$ signals the quarter-filled ordered phase. The phase is QF-SSS because 
	   $|M_b(0,0)|<|M_b(0,\pi)|,|M_b(\pi,\pi)|$. Subplot (d) displays the average SF order parameter.}
  \label{corr1_nnn}
\end{figure}

\begin{figure}
  \includegraphics[width=\linewidth]{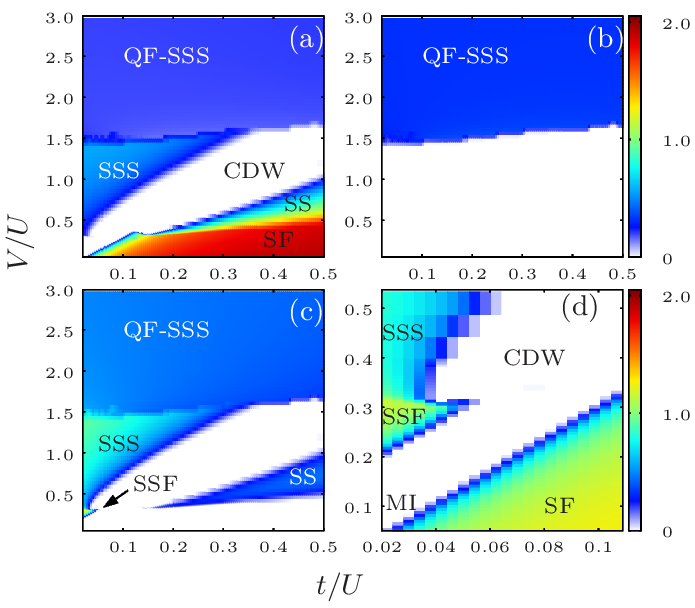}
  \caption{The Fourier transform of the single-particle correlations $M_b$ for $\rho = 2$ with NNN at (a) $(k_x,k_y) = (0,0)$, (b) $(k_x,k_y) = (0,\pi)$ and 
	   	(c) $(k_x,k_y) = (\pi,\pi)$. Unlike at unit filling, here for lower values of NN interaction a CDW-SSS transition occurs and at 
	   	$V/U \gtrsim 1.5$, the system enters into QF-SSS phase.
		The QF-SSS phase region is characterized by a finite value of $M_b(0,\pi)$ (see (b)) and of  $M_b(\pi,\pi)>M_b(0,0)$ (see (c)). {Note the existence of the SSF phase at $V/U\sim 0.25$ in the presence of the NNN density-density interactions}. Subplot (d) shows the average SF order parameter in the vicinity of the SSF phase.}
  \label{corr2_nnn}
\end{figure}

\begin{figure}
  \includegraphics[width=\linewidth]{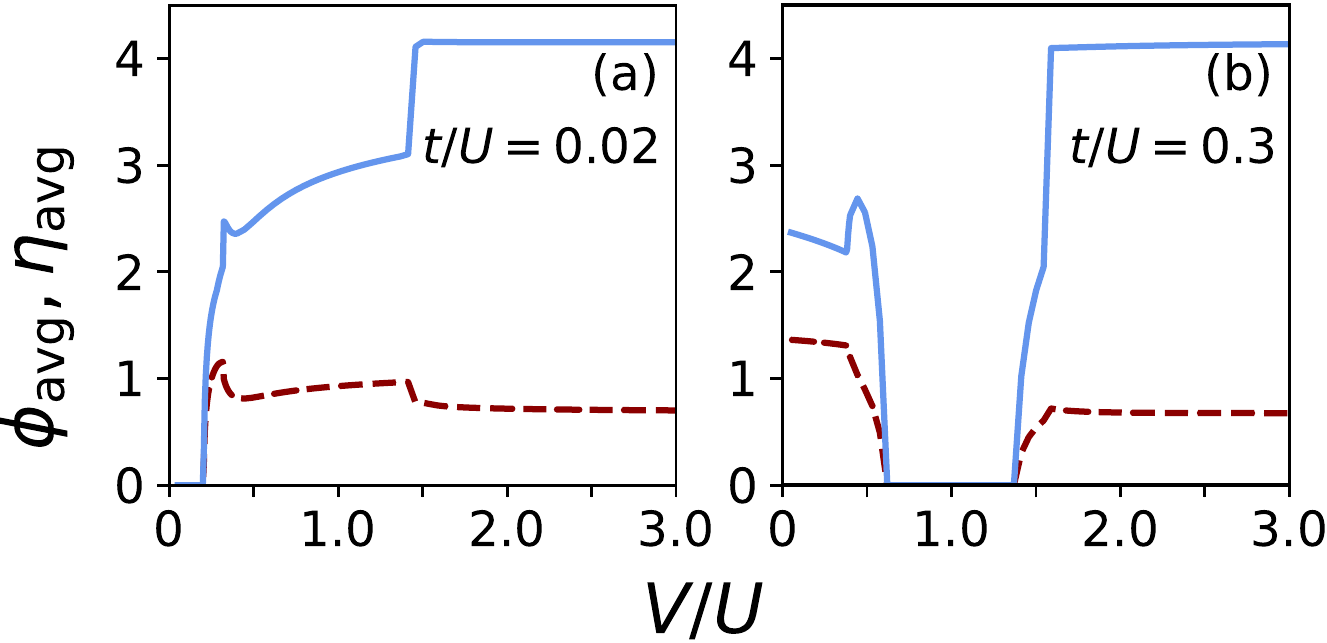}
  \caption{The average SF, Eq. (\ref{phi:av}), and density-correlated order parameter, Eq. (\ref{average_eta}),  as a function of $V/U$ for $\rho=2$ with NNN terms of the eBHM, Eq. \eqref{ebhm_h2}. Subplot (a) has been calculated for $t/U=0.02$ and 
	   (b) for $t/U=0.3$. Here, the solid blue (dashed red) line is 
	   $\eta_{\rm avg}$ ($\phi_{\rm avg}$).} 
  \label{avgop}
\end{figure}

We now include the NNN terms in our mean-field treatment. We first recall that the quantum phases of eBHM due to competition between NN and NNN repulsion at various filling have been 
studied before in the absence of the density-dependent tunnelings~\cite{ng_08,ng_10,kuno_14,heydarinasab_17}. We now include density-dependent tunneling and analyse the ground state of the model Hamiltonian in Eq.~\eqref{ebhm_mf} when the NNN terms of density-density repulsion and of correlated tunneling are included.  

We first discuss the phase diagram at density $\rho=1$. Figure~\ref{corr1_nnn} displays  the single-particle correlations and the average SF order 
parameter.  Some qualitative features are similar to the NN case, compare with Fig. \ref{corr1}. A striking difference from the NN case is the decrease of the size of 
the insulating CDW phase. Moreover, correlated hopping here gives rise to a QF-SSS phase. 

These features become more enhanced as the density is increased. Figure~\ref{corr2_nnn} shows the correlations and average SF order parameter for density $\rho=2$. With respect to the NN case (cif. Fig. \ref{corr2}) we observe that the SSS phase now disappears above a critical value $V/U\sim 1.5$ and becomes QF-SSS. The size of the QF-SSS phase is significantly larger than for unit filling. In particular, the phase boundary of CDW to QF-SSS transition weakly depends on $t/U$. These behaviours qualitatively agree with the one-dimensional phase diagram of Ref. \cite{kraus_20}. Important differences are that the SSF phase results to be smaller. Moreover, the nature of the phase, which is here QF-SSS, could not be uniquely identified in Ref. \cite{kraus_20}. 

Figure~\ref{avgop} displays {the average SF order parameter, $\phi_{\rm avg}$, and the average  density-correlated order parameter $\eta_{\rm avg}$, Eq. \eqref{average_eta}, 
as a function of $V/U$ and fixed values of $t/U$. We first consider the value $t/U = 0.02$, subplot (a). Here, at $V=0$ the phase 
is MI. As $V$ is increased, the transitions  MI-SSF-SSS take place at {$V/U\sim 0.25$} and is here signalled by a sharp increase and a local maximum of both parameters. Finally, at the transition SSS to QF-SSS $\eta_{\rm avg}$ increases whereas $\phi_{\rm avg}$ decreases. At $t/U=0.3$, the system shows SF-SS-CDW-SSS transition as $V/U$ is increased. The corresponding trends of $\eta_{\rm avg}$ and $\phi_{\rm avg}$ are visible in Fig.~\ref{avgop}(b). 

\section{Phase transitions at fixed $V/U$}
\label{pd_v}
\begin{figure}
  \includegraphics[width=\linewidth]{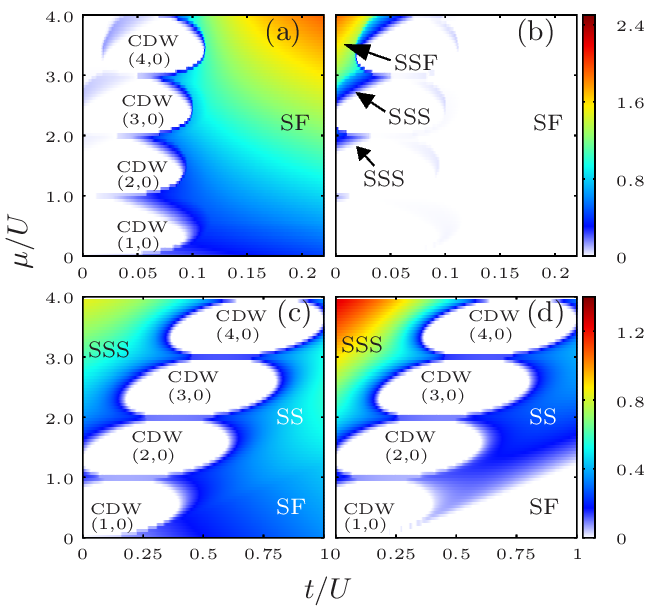}
  \caption{The off-diagonal single-particle correlations $M_{b}(0,0)$ and $M_b(\pi,\pi)$ as a function of $\mu/U$ and $t/U$ for 
	   {$V/U=0.25$} (upper panels) and {$V/U=2$} (lower panels). The left panels show the contour plot of $M_{b}(0,0)$, the right panels of $M_{b}(\pi,\pi)$. The diagram has been calculated using the SGMF method and truncating the interactions to the NN terms.}
  \label{corr_nn}
\end{figure}

We now discuss  the ground state phase diagrams in $t-\mu$ plane and for fixed values of $V/U$. For these values the NN repulsion energy becomes comparable to the on-site interaction energy and the insulating phases are  CDW~\cite{iskin_11}. At {$4V=U$}, in particular, the MI phase with $n_0$ boson per site becomes degenerate 
with the CDW with occupancies $(2n_0,0)$. Moreover, the CDW phases with $(n_0+1,n_0)$ and with  $(2n_0+1,0)$ are degenerate~\cite{iskin_11}. 
We vary $\mu$ from $0$ up to $4U$ in order to include the density $\rho=2$. In the $t-\mu$ plane the line of destructive interference, Eq. \eqref{T:t}, moves to higher $t$ values as $\mu$ is increased, since the density $\rho=\rho(\mu)$ increases monotonically with $\mu$.

\subsection{Results using SGMF}

Figure \ref{corr_nn} displays the phase correlation functions $M_b(0,0)$ and $M_b(\pi,\pi)$ for {$V/U = 0.25$} (upper panels) and {$V/U = 2$} (lower panels). The phase diagram is separated into two regions by the sequence of insulating CDW lobes, which are localized along the values of $\rho$ and $t$ fulfilling Eq. \eqref{T:t}.  On the left side, where correlated hopping dominates, the phase is SSS. The size of this region increases with $V$, and thus with the value of $|T|$. The SSF phase is observed only for {$V/U=0.25$} and at sufficiently high values of $\mu$, here at $\mu\geqslant 3$, where $\rho\approx 2$. Here, a direct SSF-CDW transition is observed in a small parameter region at $\mu\approx3.2$. On the right side single-particle hopping is responsible for the emergence of the SS phases. At higher $t$, the system enters into the SF phase. At {$V/U=2$} we observe the SS-SF phase  boundary, which varies linearly as a function of $t$. This feature is consistent with the findings of quantum Monte Carlo studies of 2D eBHM~\cite{ohgoe_12}. 
{It is important to note that the choice of boundary conditions strongly affects the phases and their transitions. As mentioned 
previously, we consider periodic boundary conditions, while the use of open boundary conditions lead to nonuniform densities at the edges 
which hinders the identification of quantum phases.}

\subsection{Comparison between SGMF and CGMF}

\begin{figure}
  \includegraphics[width=\linewidth]{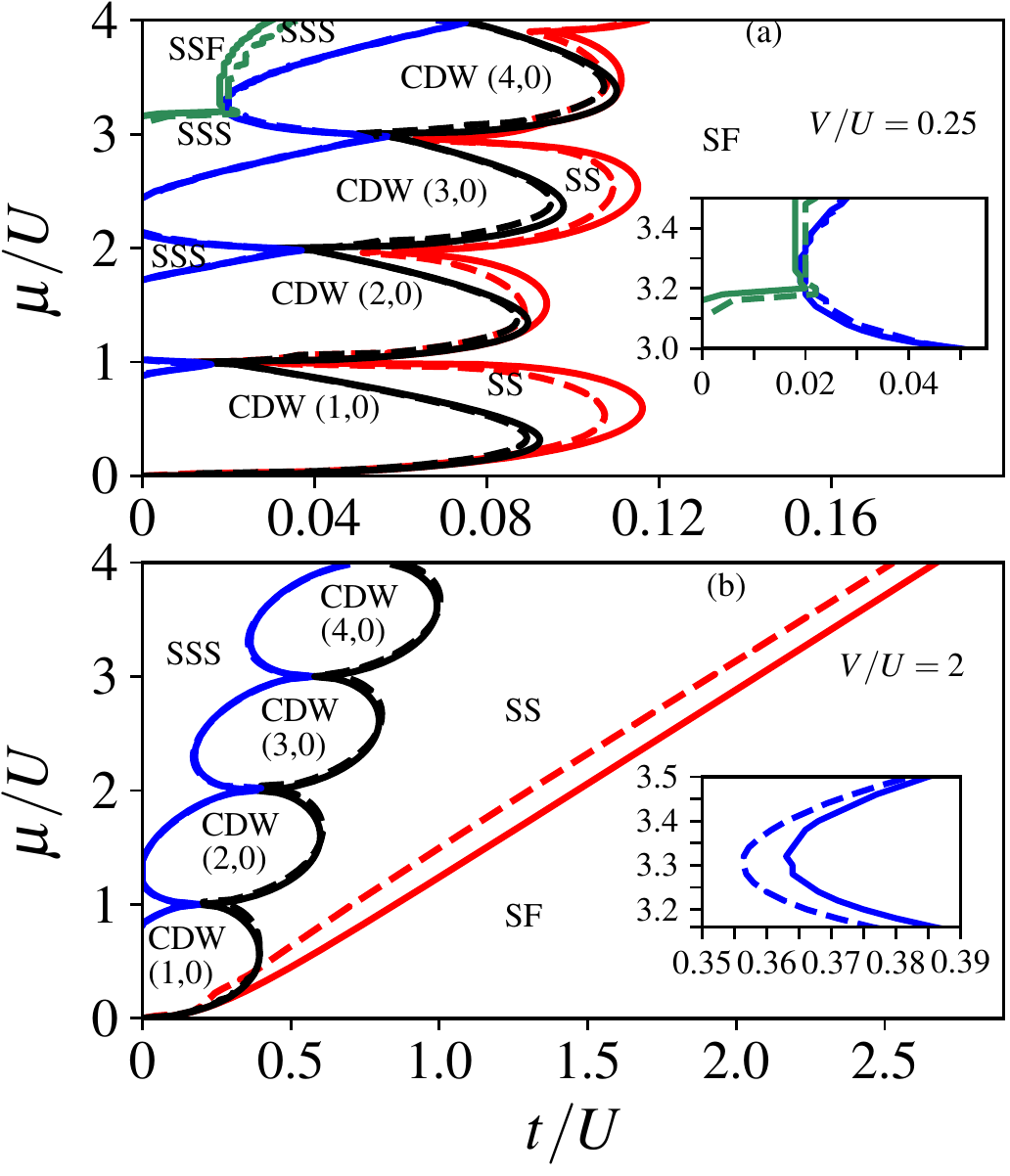}
	\caption{Ground state phase diagram of {Eq.~(\ref{ham})} with NN 
	   interactions and density-dependent tunneling. The solid and dashed lines are phase boundaries  
	   obtained using the SGMF and CGMF methods, respectively, for simulating {Eq.~(\ref{ham})}. The CGMF method 
	   uses $2\times 2$ clusters. The CDW phases are indicated by their
	   sublattice occupancies $(n_a,n_b)$. The insets zoom into regions of the phase diagram where CGMF predicts an increase of the size of the SSF phase (a) and a decrease of the size of the SSS phase (b) with respect to the SGMF predictions.}
  \label{pd_nn}
\end{figure}

We explore  the 
effects of quantum fluctuations and the intersite correlations on the 
quantum phase transitions by means of the CGMF method using $2\times 2$ cluster. We restrict here to the NN case. Figure \ref{pd_nn} shows the ground state phase diagram of our model in the $t-\mu$ plane for {$V/U=0.25$} and {$V/U=2$}. 
To illustrate the differences, the boundaries between various phase transitions 
computed with the SGMF and CGMF methods are shown. For {$V/U=0.25$}, 
Fig.\ref{pd_nn}(a), 
the CGMF method predicts smaller domains of the $(n_0,0)$ CDW lobes. For instance, the phase boundary  separating the CDW $(1,0)$ from the SS phase is at $t_c/U\approx 0.093$ for SGMF, and it is decreased to 
$t_c/U \approx 0.089$ with the CGMF. A similar trend of the phase boundary $t_c/U$ for the CDW-SS transition was reported in 
\cite{suthar_19} for {$V/U = 0.2$}. 
It is to be noted  that the size of the SSF phase increases with the CGMF method. 
This is visible in the inset of Fig.\ref{pd_nn}(a). At $\mu/U = 3.8$, the 
SSF phase persists upto $t/U \approx 0.023$ using SGMF, and with CGMF this 
is modified to $t/U \approx 0.03$. Our computations show that CGMF predicts an increase of the size of the SSF phase while the SSS-CDW phase boundary remains unaffected. This results in a reduction of the SSS phase domain. 
Thus, the quantum fluctuations captured by the CGMF tend to correct the SGMF predictions by increasing the size of the uniform density phases like the SSF and the SF, while the domains of the structured density phases like the 
CDW, SSS and SS are tendentially decreased. As with the {$V/U=0.25$} case, the CGMF results show a decrease in the domain of the SS phase in the 
phase diagram with {$V/U=2$}, from the inset in Fig.\ref{pd_nn}(b). On the contrary, the domain of the CDW phase remains unaffected.


\section{Conclusions}
\label{conc}
 We have studied the zero temperature phase diagram of dipolar bosons in 2D optical lattice using a mean-field approach
and investigated the effects of the interplay between single-particle hopping and correlated tunnelings when these can destructively interfere. 
The mean-field study, both by means of site-decoupled and  cluster Gutzwiller approach, confirms the findings of the 1D phase diagrams for density $\rho=2$ and obtained using DMRG \cite{kraus_20}. Moreover, it extends them to the grandcanonical ensemble: The interference cuts the phase diagram into two topologically different superfluid phases. In particular, there where correlated hopping dominates the SF order parameter is staggered. The comparison of the domains of SSS and SSF at fixed nearest neighbor interactions exhibits the suppression of SSS phase with the enhancement of SSF phase when the  inter-site correlations are accounted for. When next-nearest neighbour terms are included, a quarter-filled supersolidity appears at large dipolar strengths. 
The present study has been performed for experimentally consistent parameters and can serve as a guide for a possible observation of staggered quantum phases of dipolar bosons in 2D optical lattices.
{The recent experimental realization of extended Bose-Hubbard model for ultracold gas with strong dipolar interaction is an ideal setup where the quantum states here studied could be observed~\cite{baier_16}.}

Finally, we have verified that the quantum phases and phase diagrams reported in the present work are robust to the system size, by repeating the calculations for $30x30$ lattice site system.

\acknowledgments
We thank K. Biedr\'on for discussions and acknowledge the support of PL-Grid Infrastructure, Poland and Vikram-100, the 100TFLOP HPC Cluster at
Physical Research Laboratory, Ahmedabad, India. K.S. and J.Z. acknowledge the support by  National Science
Centre (Poland) under project 2016/21/B/ST2/01086. R.K. and G.M. acknowledge the support by the German Research Foundation (the priority program
No. 1929 GiRyd) and by the German Ministry of Education and Research (BMBF) via the QuantERA project NAQUAS. Project NAQUAS has received funding 
from the QuantERA ERA-NET Cofund in Quantum Technologies implemented within the European Union's Horizon 2020 program.

\section*{Appendix}
We discuss here the ground-state phase diagram of the model considered in the absence of the density dependent tunnelings 
($T$ is put to zero by force).
In Fig.~\ref{corr_no_ddt}, the single-particle correlations $M_b$ at $(0,0)$ and $(\pi,\pi)$ are 
shown for the average densities $\rho=1$ and $\rho=2$. We first discuss the quantum phase transitions 
at $\rho=1$. The $M_b(0,0)$ which is a measure of off-diagonal long-range order, has a finite value for compressible 
SF and SS phases whereas it remains zero for insulating MI and CDW phases. At low NN interaction
{$V/U \ll 0.25$}, there is MI-SF transition where the critical hopping $t_c$ is independent on $V$.
When NN interaction is comparable or overcomes to the onsite interaction {$V/U \geqslant 0.25$}, the transition between 
density modulated quantum phases CDW-SS occurs. The $t_c$ of CDW-SS transition increases as a function 
of $V$~\cite{kimura_11}. The finite value of $M_b(\pi,\pi)$ for SS phase clearly demarcates it from SF and 
CDW phases, as evident from Fig.~\ref{corr_no_ddt}(b). At $t/U=0.05$, the MI-CDW transition of $\rho=1$ as a function 
of $V$ is consistent to the previous quantum Monte Carlo study~\cite{sengupta_05}. At $\rho=2$, the qualitative features of various 
phase transitions remain similar to $\rho=1$ case, however quantitatively the critical hopping varies [Fig.~\ref{corr_no_ddt}(c,d)].
\begin{figure}
  \includegraphics[width=\linewidth]{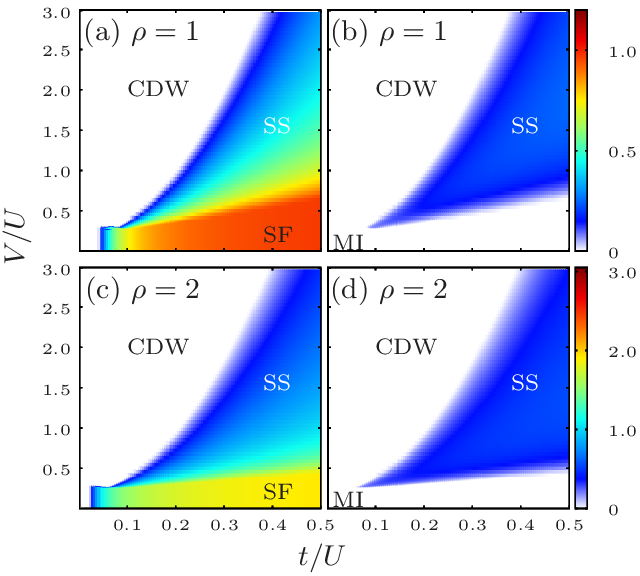}
  \caption{The single-particle correlations with NN interaction of eBHM and $T=0$.
	   The correlations for $\rho=1$ are shown at (a) $(k_x,k_y) = (0,0)$ and (b) $(k_x,k_y) = (\pi,\pi)$,
	   and for $\rho=2$ these are shown at (c) $(k_x,k_y) = (0,0)$ and (d) $(k_x,k_y) = (\pi,\pi)$}
  \label{corr_no_ddt}
\end{figure}


%

\end{document}